\documentclass[11pt,a4paper]{article}
\pdfoutput=1

\usepackage{jcappub} 

\usepackage{txfonts}

\def\lapprox{\mathrel{\mathop  {\hbox{\lower0.5ex\hbox{$\sim$}
\kern-1.1em\lower-0.7ex\hbox{$<$}}}}}
\def\gapprox{\mathrel{\mathop  {\hbox{\lower0.5ex\hbox{$\sim$}
\kern-1.1em\lower-0.7ex\hbox{$>$}}}}}

\title{ The cosmological \boldmath $^{7}{\rm Li}$ problem from a nuclear 
physics perspective}

\author[a]{C.~Broggini,}
\author[a]{L.~Canton,} 
\author[b,d]{G.~Fiorentini}
\author[c,e] {and 
F.~L.~ Villante}

\affiliation[a]{Istituto Nazionale di Fisica Nucleare (INFN), 
Sezione di Padova, Padova I-35131, Italy}
\affiliation[b]{Istituto Nazionale di Fisica Nucleare (INFN), 
Laboratori Nazionali di Legnaro (LNL), Legnaro (PD) I- 35020, Italy}
\affiliation[c]{Istituto Nazionale di Fisica Nucleare (INFN), 
Laboratori Nazionali del Gran Sasso (LNGS), Assergi (AQ) I-67010, Italy}
\affiliation[d]{Dipartimento di Fisica, Universit\`a di Ferrara, Ferrara I-44100, Italy}
\affiliation[e]{Dipartimento di Fisica, Universit\`a di L'Aquila, L'Aquila I-67100, Italy}

\emailAdd{carlo.broggini@pd.infn.it}
\emailAdd{luciano.canton@pd.infn.it}
\emailAdd{giovanni.fiorentini@fe.infn.it}
\emailAdd{francesco.villante@lngs.infn.it} 

\abstract{The primordial abundance of $^{7}{\rm Li}$ as predicted by Big Bang 
Nucleosynthesis (BBN) is more than a factor 2 larger than what has been 
observed in metal-poor halo stars.  Herein, we analyze the possibility 
that this discrepancy originates from incorrect assumptions about the nuclear 
reaction cross sections relevant for BBN. To do this, we introduce an 
efficient method to calculate the changes in the $^{7}{\rm Li}$ abundance 
produced by arbitrary (temperature dependent) modifications of the nuclear 
reaction rates.  Then, considering that $^7{\rm Li}$ is mainly produced from $^7{\rm Be}$ 
via the electron capture process $^{7}{\rm Be}+{\rm e}^-\to$$^{7}{\rm Li}+\nu_{\rm e}$,  we assess the  impact of the various channels 
of $^{7}{\rm Be}$ destruction.
Differently from previous analysis, we consider the role of unknown resonances by using 
a complete formalism which takes into account the effect of Coulomb and centrifugal barrier penetration and that does 
not rely on the use of the narrow-resonance approximation. 
As a result of this, the possibility of
a nuclear physics solution to the $^{7}{\rm Li}$ problem is significantly suppressed.
Given the present experimental and theoretical constraints, it is unlikely that the $^{7}{\rm Be} + n$ destruction rate 
is underestimated by the 2.5 factor required to solve the problem. We exclude, moreover, that
resonant destruction in the channels $^7{\rm Be}$ + $t$ and $^7{\rm Be}$ + 
$^3{\rm He}$ can explain the $^{7}{\rm Li}$ puzzle. New unknown resonances in 
$^7{\rm Be}$ + $d$ and $^7{\rm Be}$ + $\alpha$ could potentially produce 
significant effects. Recent experimental results have ruled out such a possibility for $^7{\rm Be}+d$. 
On the other hand, for the $^7{\rm Be}+\alpha$ channel
very favorable conditions are required. 
The possible existence of a partially suitable resonant level in $^{11}{\rm C}$ 
is studied in the framework of a coupled-channel model and the possibility of a direct measurement is considered.}


\begin{document}
\maketitle

\newpage

\section{Introduction}

Big Bang Nucleosynthesis (BBN) is one of the solid pillars of the standard 
cosmological model and represents the earliest event in the history of the 
universe for which confirmable predictions can be made 
(see~\cite{Nakamura:2010zzi} for a review).  The theory predicts that relevant 
abundances of light elements, namely $^{2}{\rm H}$, $^{3}{\rm He}$, 
$^{4}{\rm He}$ and $^{7}{\rm Li}$, were produced during the first minutes of 
the evolution of the universe.
Theoretical calculations of these abundances are well defined and are very 
precise. The largest uncertainties arise from the values of 
cross-sections of the relevant nuclear reactions and are at the 
level of 0.2\% for $^{4}{\rm He}$, 5\% for $^{2}{\rm H}$ and $^{3}{\rm He}$  
and 15\% for $^{7}{\rm Li}$.\footnote{
After the first evaluations~\cite{SKM}, 
theoretical uncertainties in BBN have been carefully assessed in several
papers. Monte-Carlo and semi-analytical approaches have been used. 
As examples see Refs.~\cite{serpicoetal,noi}.}

In standard BBN, the primordial abundances depend on only one free parameter, 
the present baryon-to-photon ratio
$ \eta\equiv (N_{\rm B}-N_{\overline{\rm B}})/N_{\gamma} $, which is related to 
the baryon density of the universe by 
$\Omega_{\rm B} h^{2} = 3.65\cdot 10^{7} \,\eta$. 
This quantity can be constrained with high accuracy from the observation of 
the anisotropies of the Cosmic Microwave Background (CMB). The latest 
WMAP-7 results suggest $\Omega_{\rm B} h^{2} = 0.02249\pm 0.00056$, which 
corresponds to 
$\eta_{\rm CMB} = 6.16 \pm 0.15 \times 10^{-10}$ \cite{Komatsu:2010fb}. 
If this value is accepted, then BBN is a parameter free theory which can be 
used to test the standard cosmological model and/or the chemical evolution of 
the universe.

Comparison of theoretical predictions with observational data is not 
straightforward.  Data are subject to poorly known evolutionary effects and 
there are systematic errors.  Even so, the agreement between the 
predicted primordial abundances of $^{2}{\rm H}$ and $^{4}{\rm He}$ and 
the values inferred from observations is non-trivial. However, the
situation is much more complicated for $^7$Li.  Using $\eta= \eta_{\rm CMB}$, 
the predicted primordial $^{7}{\rm Li}$ abundance is \cite{Cyburt:2008kw}
\begin{equation}
{\rm (Li/H)}_{\rm BBN}\simeq (5.1 ^{+0.7}_{-0.6} ) \times 10^{-10} .
\end{equation}
This is a factor $\sim 3$ larger than that inferred by observing the 
so-called 'Spite Plateau' in the $^7$Li abundance of metal-poor halo
stars, which has been given~\cite{Nakamura:2010zzi} as
\begin{equation}
{\rm (Li/H)}_{\rm obs}\simeq (1.7  \pm 0.06 \pm 0.44  ) \times 10^{-10} .
\end{equation}
The quoted errors take into account the dispersion of the various observational 
determinations.
Moreover, it is considered that Lithium in Pop II stars can be destroyed as a 
consequence of mixing of the outer layers with the hotter interior. This process 
can be constrained by the absence of significant scatter in Li versus Fe in the 
Spite Plateau \cite{Ryan1999}.

 The abundance of $^7{\rm Li}$ is a central unresolved issue in 
BBN~\cite{Cyburt:2008kw} about which there has been recent concern~\cite{Li7obs, Ryan1999, Asplund2006}   
regarding erroneous evaluation of nuclear reaction rates 
responsible for $^7{\rm Li}$ production. 
At $\eta=\eta_{\rm CMB}$, $^7$Li is mainly produced from $^7$Be 
via the electron capture process 
${\rm e}^{-}+{} ^7{\rm Be}\to{} ^7{\rm Li}+\nu_{\rm e}$.
Thus nuclear reactions producing and destroying $^{7}{\rm Be}$ must be
considered. The leading processes,
$^{3}{\rm He}(\alpha,\gamma)^{7}{\rm Be}$ and $^{7}{\rm Be}(n,p)^{7}{\rm
Li}$, have been well studied and the cross sections are known
to a few percent accuracy~\cite{serpicoetal}. 
In \cite{Coc:2003ce}, it was noted that an increase by a factor greater than
$1000$ in the sub-dominant $^7{\rm Be}(d,p)2\alpha$ cross section could
provide the necessary suppression of $^7{\rm Li}$. This enhancement was not found in experimental data \cite{Angulo:2005mi}
but could have escaped detection if it were produced by a sufficiently narrow resonance, as suggested in \cite{Cyburt:2009cf}. 
Other possible resonant destruction channels have been considered
\cite{Chakraborty:2010zj} as well, such as the channels 
$^{7}{\rm Be}+{} ^{3}{\rm He}\to{} ^{10}{\rm C}$ and  
$^{7}{\rm Be}+t \to{} ^{10}{\rm B}$ that await experimental verification.

In this paper, we consider further the cosmological $^7{\rm Li}$ problem from
the nuclear physics perspective. To do this, in Sec.~\ref{LiResponse}, we
introduce an efficient method to calculate the response of the  $^7{\rm Li}$ 
primordial abundance to an arbitrary modification of the nuclear reaction rates.
This approach leads to an understanding, in simple physical terms, of
why it is so difficult to find a nuclear physics solution to the observed 
discrepancy. Then, in Secs.~\ref{Beneutroni} and \ref{Becharged}, 
the various possible $^{7}{\rm Be}$ destruction channels are considered,
including possible new resonances. This has been suggested recently also 
in Ref.~\cite{Chakraborty:2010zj}, but here we use a more refined 
description of the nuclear processes than in previous investigations. 
In particular, we do not assume the narrow-resonance approximation 
and we include the effect of the Coulomb and centrifugal barrier penetration
 in our parametrisation of resonating cross sections. As a result, 
the parameter space for a nuclear physics solution of the $^{7}{\rm Li}$ puzzle 
is considerably reduced.
Our conclusions are summarised in Sec.\ref{Conclusions}.

\section{
The \boldmath $^{7}{\rm Li}$ response to nuclear reaction rate modifications}
\label{LiResponse}

It is useful to briefly review the light element production mechanism in the 
early universe.  The abundance of a generic element $i$ in the early
universe, $Y_{i} = n_{i}/n_{\rm B}$ where $n_{\rm B}$ is the baryon 
number density, evolves according to the rate equations,
\begin{equation}
\frac{d Y_{i}}{d t} = n_{\rm B} \left[ \sum_{j,k} Y_{j}\, Y_{k} \, 
\langle \sigma_{jk} \, v \rangle_T - \sum_{l} Y_{i} \, Y_{l} \, 
\langle \sigma_{il} \, v \rangle_T \right] .
\end{equation}
The sums include the relevant production and destruction reactions and
$\langle \sigma_{ij} \, v \rangle_T $ are the thermally averaged cross sections. 
It is known~\cite{Esmailzadeh} that a good approximation 
is obtained by studying the quasi-fixed point of the above equations, {\it viz.} 
\begin{equation}
Y_{i} \sim \left. \frac{C_{i}}{D_{i}}\right|_{T=T_{i,\rm f}} ,
\label{rate_equations}
\end{equation}
where $C_{i}$ and $D_{i}$ are the total rate of creation and destruction of the 
i-element, given by
\begin{equation}
C_{i} = n_{\rm B}\;\sum_{j,k} Y_{j} \, Y_{k}   \, \langle \sigma_{jk} 
\, v \rangle_T ,
\end{equation}
and
\begin{equation}
D_{i} = n_{\rm B}\sum_{l} Y_{l} \, \langle \sigma_{il} \, v \rangle_T .
\end{equation}
$T_{i, \rm f}$ is the freeze-out temperature for the i-element, namely
the temperature below which the rates $C_{i}$ and $D_{i}$ become smaller 
that the Hubble expansion rate (see Ref.~\cite{Esmailzadeh} for details).
 
\begin{figure}[t]
\par
\begin{center}
\includegraphics[width=7.5cm,angle=0]{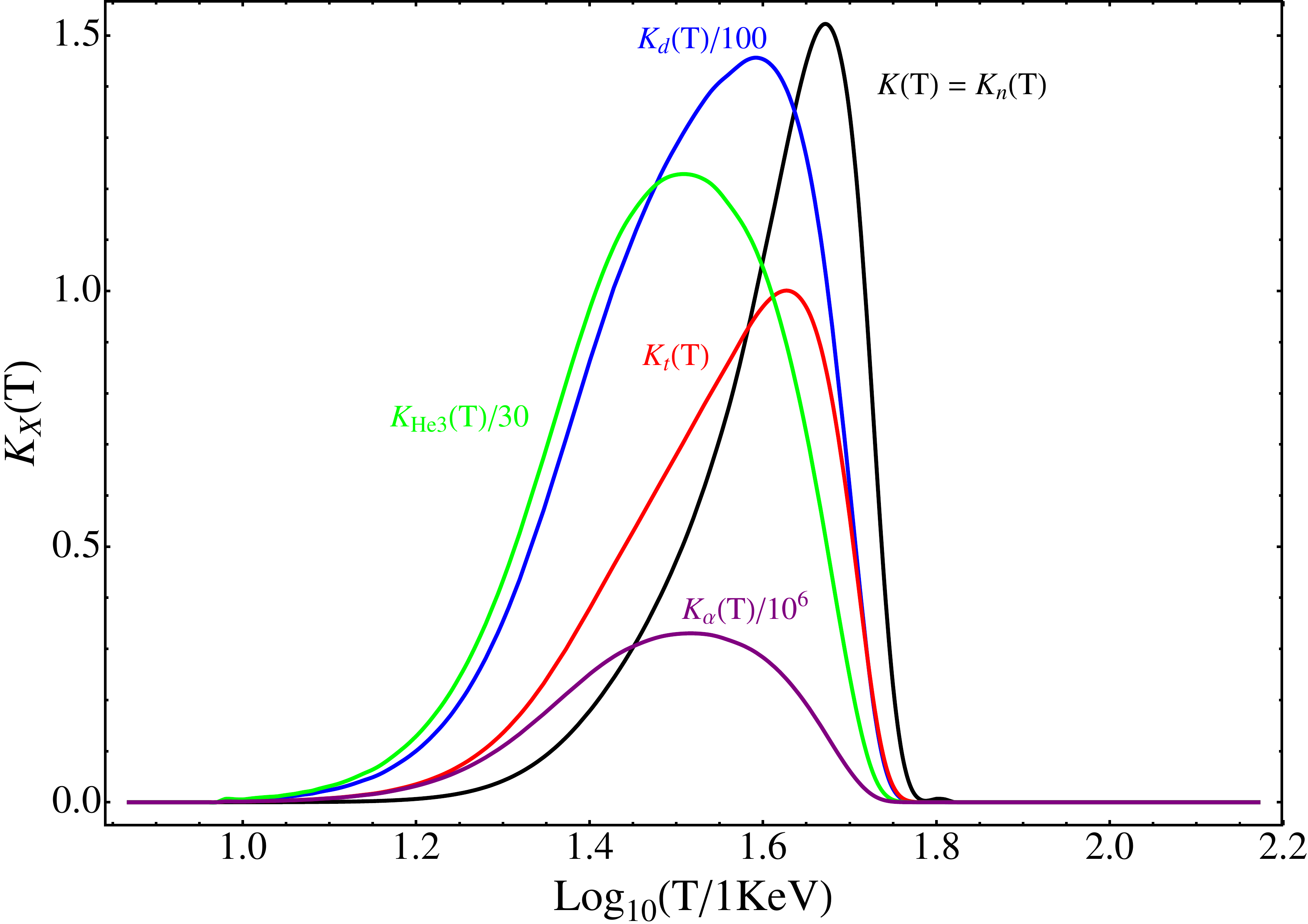}
\includegraphics[width=7.5cm,angle=0]{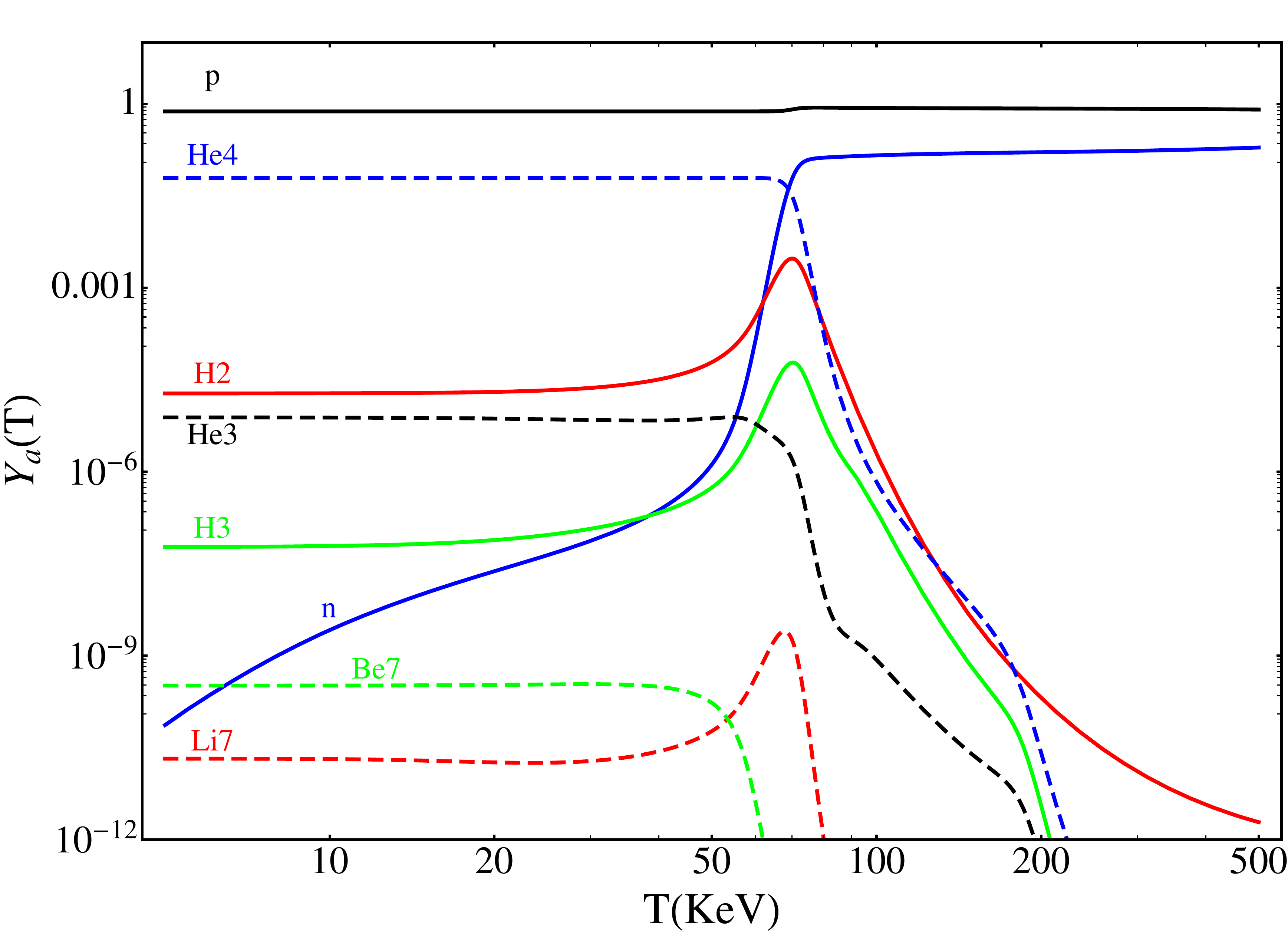}
\end{center}
\par
\vspace{-5mm} \caption{\em 
{\protect\small Left Panel: The kernels $K_{a}(T)$ defined in 
Eqs.(\ref{inverse},\ref{inverse2}).
Right Panel: The evolution of light element abundances $Y_{\rm i} = n_{\rm i}/n_{\rm B}$ as a function of temperature, assuming $\eta=\eta_{\rm CMB}$ in standard BBN.}}
\label{Fig1}
\end{figure}

At $\eta = \eta_{\rm CMB} \simeq 6\times 10^{-10}$, 
$^7$Li is mainly produced from $^7{\rm Be}$ that undergoes at late times 
($i.e.$ long after the $^7{\rm Be}$ freeze-out) the electron capture process 
${\rm e}^{-}+{} ^7{\rm Be}\to{}  ^7{\rm Li} + \nu_{\rm e}$, so that we have
\begin{equation}
\label{esmail}
Y_{\rm Li} \sim Y_{\rm Be}  \sim  \left. 
\frac{C_{\rm Be}}{D_{\rm Be}}\right|_{T=T_{\rm Be, f}} ,
\end{equation}
where $C_{\rm Be}$ and $D_{\rm Be}$ are the total $^{7}{\rm Be}$ production and 
destruction rates.
The dominant $^{7}{\rm Be}$ production mechanism is through the capture
reaction, $^3{\rm He}(\alpha,\gamma)^7{\rm Be}$; a reaction  that
has been studied in detail both experimentally~\cite{Luna} and theoretically~\cite{Can-Lev}. 
The cross section is known to $\sim 3\%$ uncertainty. 
The dominant $^{7}{\rm Be}$ destruction channel is the process
 $^7{\rm Be}(n,p)^7{\rm Li}$; 
the cross section of which now is known to $\sim 1\%$ accuracy 
as we discuss in the next section.
As these leading processes have been well studied, a sizeable reduction of the 
$^{7}{\rm Li}$ predicted abundance occurs only if large increases of the 
sub-dominant $^{7}{\rm Be}$ destruction channels are allowed as may be the
case if new, so far unknown, resonances become influential. 
To study this possibility, we introduce a simple formalism to describe
the response of $^{7}{\rm Li}$ to a generic (temperature dependent) modification 
of the nuclear reaction rates.
Motivated by Eq.(\ref{esmail}), we assume that a linear relation exists
between the {\em inverse} $^7$Li abundance, 
$X_{\rm Li} \equiv 1/Y_{\rm Li}$, and the total $^7$Be destruction rate. This 
relation can be expressed in general terms as\footnote{Here and in the following, 
the notation $\overline{Q}$ refers to the standard value for the generic quantity 
$Q$.},
\begin{equation}
\delta X_{\rm Li} = \int \frac{dT}{T} \; K(T) \; \delta D_{\rm Be}(T) ,
\label{inverse}
\end{equation} 
where
\begin{equation}
\delta X_{\rm Li} = \frac{X_{\rm Li}}{\overline{X}_{\rm Li}} - 1 ,
\end{equation}
and
\begin{equation}
\delta D_{\rm Be}(T) = \frac{D_{\rm Be}(T)}{\overline{D}_{\rm Be}(T)} -1 .
\end{equation}
The integral kernel $K(T)$ has been evaluated numerically by considering the 
effects of localised (in temperature) increases of the reaction rate 
$D_{\rm Be}(T)$. Our results are shown by the black solid line in the left 
panel of Fig.~\ref{Fig1}. 
We checked numerically that Eq.~(\ref{inverse}) describes 
accurately large  variations of the $^{7}{\rm Li}$ abundance  (up to a factor $\sim 2$) and, thus,
it is adequate for our purposes. 
The kernel $K(T)$ is peaked at $\sim 50$ keV; roughly corresponding to the 
$^7{\rm Be}$ freeze-out temperature $T_{\rm Be, f}$. The area under the curve 
is equal to $\simeq 0.7$ which indicates that the total destruction rate of
$^7$Be should be increased by a factor $\sim 2.5$ to obtain a factor 2
reduction in the abundance of $^7$Li.

We can use Eq.~(\ref{inverse}) to assess the sensitivity of the abundance
of $^{7}{\rm Li}$ with respect to a specific reaction channel.
The total $^7$Be reaction rate is given by
\begin{equation}
D_{\rm Be}(T) =  n_{\rm B}  \, \sum_{a} Y_{a}(T) \, \langle \sigma_{a} v 
\rangle_T ,
\end{equation}
where $\sigma_{a}$ is the cross section of the reaction $^{7}{\rm Be}+ a$ and 
$Y_{a}$ represents the elemental abundance of the $a$ nuclei.
In standard BBN, the dominant contribution is provided by the
$^{7}{\rm Be}(n,p)^{7}{\rm Li}$ reaction; accounting for about $\sim 97 \%$ of 
the total $^7{\rm Be}$ destruction rate. The standard rate 
$\overline{D}_{\rm Be}(T)$ can then be set with a few percent accuracy by, 
\begin{equation}
\overline{D}_{\rm Be}(T) \simeq n_{\rm B}  \, \overline{Y}_{\rm n}(T) \, 
\langle \overline{\sigma}_{\rm np} v \rangle_T ,
\end{equation}
where $Y_{\rm n}(T)$ is the neutron abundance and $\sigma_{\rm np}$ is the 
cross section of $^{7}{\rm Be}(n,p)^{7}{\rm Li}$. 
Sub-dominant reaction channels can provide a non-negligible contribution 
only if there is a large increase of their assumed cross section values. 
The fractional enhancement of $D_{\rm Be}(T)$ due to a generic 
$^{7}{\rm Be}+a$ process can be evaluated from,
\begin{equation}
\delta D_{{\rm Be}, a}(T)  = 
\frac{\overline{Y}_{a}(T)}{\overline{Y}_{\rm n}(T)}\frac{\langle \sigma_{a} 
v \rangle_T}{\langle \overline{\sigma}_{\rm np} v \rangle_T} ,
\end{equation}
under the reasonable assumption\footnote{
As we see in the right panel of Fig.~\ref{Fig1}, the abundance of Beryllium is 
much lower than the abundances of $d$, $t$, $^3{\rm He}$ and $^{4}{\rm He}$. 
This implies that a tiny fraction of these elements can potentially produce
a very large depletion of Beryllium nuclei.} that 
the inclusion of a new channel for $^{7}{\rm Be}$ destruction
does not alter the abundance of the $a$ nuclei. 
In the right panel of Fig.~\ref{Fig1},  we show the light element abundances,
 $\overline{Y}_{a}(T)$, calculated assuming $\eta = \eta_{\rm CMB}$ in standard 
BBN.  We see that $d$, $^3{\rm He}$ and $^{4}{\rm He}$ have abundances larger or 
comparable to that of neutrons at the temperature $T\sim 50$ keV relevant for 
$^7$Be synthesis.  Thus reactions involving these nuclei could provide a 
non-negligible contribution to $D_{\rm Be}(T)$
even if their cross sections are lower than that of 
$^{7}{\rm Be}(n,p)^{7}{\rm Li}$.
This point can be expressed quantitatively by rewriting Eq.~(\ref{inverse})
as
\begin{equation}
\delta X_{\rm Li} = \sum_{a} \int \frac{dT}{T} \; K_{a}(T) \;  
\frac{\langle \sigma_{a} v \rangle_T}{\langle \overline{\sigma}_{\rm np} 
v \rangle_T} ,
\label{inverse2}
\end{equation}
where
\begin{equation}
\label{kernels}
K_{a}(T) = K(T) \frac{\overline{Y}_{a}(T)}{\overline{Y}_{\rm n}(T)} .
\end{equation}
The kernels $K_{a}(T)$ are shown in the left panel of Fig.~\ref{Fig1} 
for the cases 
$a = n, \; d, \; t, \; ^{3}{\rm He}, \; ^{4}{\rm He}$
and can be used to quantify the requirements for a nuclear physics solution
of the cosmic $^7{\rm Li}$ problem. To be more consistent with observed 
data, a reduction of the $^{7}{\rm Li}$ abundance by a factor 2 or more is
required and that corresponds to $\delta X_{\rm Li} \ge 1$.
To obtain this, the ratios $R_a \equiv \langle \sigma_{a} v \rangle_T / 
\langle \overline{\sigma}_{\rm np} v \rangle_T$
at temperatures $T\simeq 10 - 60 \, {\rm keV}$ should be,
$R_{\rm n} \ge 1.5$ for additional reactions in the $^7{\rm Be}+n$ channel,
$R_d \ge 0.01$ for reactions in the $^7{\rm Be}+d$ channel,
$R_t \ge 1.5$ for  reactions in the $^7{\rm Be}+t$ channel, 
$R_{\rm He3} \ge 0.03$ for  reactions in the $^7{\rm Be}+{}^3{\rm He}$
channel, and
$R_{\rm He4} \ge 4\times 10^{-6}$ for  reactions in the 
$^7{\rm Be}+{}^4{\rm He}$ channel.
In the next section, we explore these possibilities on the basis of 
general nuclear physics arguments.

\section{Treatment of nuclear cross-sections}
\label{AppA}

Except for spin statistical factors, the partial reaction cross section for a 
generic process, $^7{\rm Be} + a$, cannot be larger than
$\sigma_{\rm max}= (2l+1)\, \pi \, \lambdabar^2$
where $l$ is the orbital angular momentum of the channel considered, 
$\lambdabar = 1 / k$, and $k$ is the momentum in the center of mass (CM).
The relation can be rewritten as
\begin{equation}
\sigma_{\rm max} = (2l+1)\, \pi \, \lambdabar^2
= (2l+1)\, \frac{\pi }{2 \mu \, E}
\end{equation}
where $E$ is the CM energy and $\mu$ is the reduced mass of the colliding 
nuclei.

 For low-energy reactions involving charged nuclei and/or a non-vanishing 
angular momentum, the cross section is suppressed due 
quantum tunnelling through the Coulomb and/or centrifugal barrier. 
Modelling the nuclear interaction potential by a square well with a radius $R$, 
the partial cross section for the formation of a compound system $C$ in the 
process $^{7}{\rm Be} + a \rightarrow (C) $ can be expressed as,
\begin{equation}
\sigma_{\rm C} = \sigma_{\rm max}  \; T_{ l} .
\end{equation}
The factor $T_{\rm l}$ represents the transmission coefficient for the  channel
considered. In the low-energy limit, $k\ll K$, it can be calculated from,
\begin{equation}
T_{l}  = \frac{4 k}{K}\;v_l .
\label{TranCoeffLE}
\end{equation}
$K$ is the relative momentum of the particles inside the compound system. 
The functions $v_l$ are known as penetration factors~\cite{Blatt}. 

For uncharged particles, the penetration factors $v_l$ are given by 
\begin{equation}
v_{l} \equiv \frac{1}{G^2_{l}(R)+F^2_{l}(R)}
\label{PF}
\end{equation}
where $F_{l}(R)$ and $G_{l}(R)$ are the regular and irregular solutions of 
the Schr\"odinger radial equation, which have been tabulated in Ref.~\cite{Abra}. 
Feshbach and Lax in 1948 also tabulated the associated
functions $v'_l$. For the lowest angular momenta, 
\begin{eqnarray}
\nonumber
v_{0} &=& 1 \hspace{5.0 cm} v'_{0} =1\\
\nonumber
v_{1} &=& \frac{x^2}{1+x^2} \hspace{4.1 cm} v'_{1} =
\frac{1}{x^2}+\left(1-\frac{1}{x^2}\right)^2 ,
\end{eqnarray}
where $x \equiv k \, R = \sqrt{2\mu\, E} \; R$.
The transmission coefficient for reactions involving neutrons can be calculated 
exactly. One obtains
\begin{equation}
T_{l} = \frac{4\,  x\,  X \, v_{l}}{X^2 + (2 \, x \; X + x^2\, v'_{l})\,
v_{l}} ,
\label{TranCoeffNeutr}
\end{equation}
where $X \equiv  K \, R$. Eq.~(\ref{TranCoeffNeutr})
coincides with Eq.~(\ref{TranCoeffLE}) in the low-energy limit.

For charged nuclei, one has to rely on numerical calculations. An  approximate 
expression for the penetration factors $v_l$ can be obtained by using the WKB 
approximation. For  the collision energy $E$ lower that the height of the 
potential barrier,
\begin{equation}
v_{l} = \frac{k_{l}(R)}{k} \; \exp\left[-2\int^{r_0}_{R} k_l(r)\, dr
\right] ,
\label{WKB}
\end{equation}
where $r_0$ is the classical distance of closest approach while $k_l(r)$ is given 
by
\begin{equation}
k_{l}(r) = \sqrt{ 2\mu\,U_{l}(r)-k^2} ,
\end{equation}
with
\begin{equation}
U_{l}(r) = \frac{Z_{a}Z_{X} e^2}{r}+ \frac{l(l+1)}{2\mu r^2}  .
\end{equation}
Eq.~(\ref{WKB}), however, does not produce accurate results when the collision 
energy is close to the height of the potential barrier and then
exact expressions for the penetration factors $v_{l}$ have to be used.  
When the Coulomb interaction is taken into account, Eq.~(\ref{PF}) is still 
valid, but now $F_{l}(R)$ and $G_{l}(R)$ are Coulomb Functions, 
namely the regular and irregular solutions of the Schr{\"o}dinger radial 
equation that include the Coulomb potential. 
Such functions can be evaluated numerically with standard numerical techniques.

In the presence of an isolated resonance, the cross section for a generic process
$^7{\rm Be}+ a \rightarrow  C^* \rightarrow b +Y$ can be described by the 
Breit-Wigner formula,
\begin{equation}
\sigma = 
\frac{\pi\,\omega}{2 \mu \, E} 
\frac{\Gamma_{\rm in}\Gamma_{\rm out}}{(E-E_{r})^2+\Gamma^2_{\rm tot}/4} ,
\label{BWA}
\end{equation}
where $E_r$ is the resonance energy, $\Gamma_{\rm in}$ is the width of the 
entrance channel, $\Gamma_{\rm out}$ is the width for the exit channel 
and $\Gamma_{\rm tot} = \Gamma_{\rm in}+\Gamma_{\rm out} +{\dots}$ is the 
total resonance width. 
The first factor in the right hand side of Eq.~(\ref{BWA}) is an upper limit 
for the cross section. Basically it coincides with $\sigma_{\rm max}$ apart 
from the factor,
\begin{equation}
\omega = \frac{2 J_{C^*}+1}{(2 J_{a}+1)(2 J_7 +1)} ,
\label{omega}
\end{equation}
that takes into account the angular momenta $J_{a}$ and $J_7$ of the
colliding nuclei and the angular momentum $J_{C^*}$ of the 
excited state in the compound nucleus.
The resonance width $\Gamma_{\rm in}$ (and $\Gamma_{\rm out}$) can be expressed 
as the product,
\begin{equation}
\Gamma_{\rm in} = 2 P_{l}(E,R) \; \gamma^2_{\rm in} ,
\label{gammain}
\end{equation}
where the functions $P_{l}(E,R)$ describe the Coulomb and centrifugal barrier 
penetrations and are related to the penetration factors $\nu_l$ 
by\footnote{Note that in Ref.~\cite{Iliadis}, the quantities $P_{l}(E,R)$
themselves are referred to as the penetration factors.},
\begin{equation}
P_{l}(E,R)\equiv k R\, \nu_{l} .
\label{Pl}
\end{equation}
The reduced width, $\gamma^2_{\rm in}$, incorporates all the unknown properties 
of the nuclear interior.  This has to be smaller that the Wigner limiting
width, $\gamma^2_{\rm W}$ that is given by \cite{wigner},
\begin{equation}
\gamma^2_{\rm in} \le  \gamma^2_{\rm W} = \frac{3 }{2 \mu R^2} ~.
\label{gammaW}
\end{equation}

\section{The \boldmath $^7{\rm Be} + n$ channel}
\label{Beneutroni}

At $T\sim T_{\rm Be, f}\simeq 50\, {\rm keV}$, the $^7{\rm Be}(n,p)^7{\rm Li}$ 
reaction  accounts for $\sim 97\%$ of the total $^7$Be destruction rate. 
This process has been very well studied. Experimental data have been obtained 
either from direct measurements or from the $^{7}{\rm Li}(p, n)^{7}{\rm Be}$ 
inverse reaction (see Ref.~\cite{adahchour} and references contained therein 
for details).  The cross section for this reaction near threshold is strongly 
enhanced by a $2^{-}$ resonance at $E_x = 18.91\;{\rm MeV}$.\footnote{
The entrance energy of the $^7{\rm Be}+n$ channel with respect to the 
$^8{\rm Be}$ ground state is $E_{\rm in} = 18.8997 \; {\rm MeV}$. 
Thus the listed resonances correspond to a collision kinetic energy equal to 
$E_{r} = E_x - E_{\rm in} = 0.01,\; 0.17, \; 0.34 \; {\rm and} \; 2.6 \; 
{\rm MeV}$, respectively.}
In addition, the data show evidence for two peaks that correspond to the 
(unresolved) $3^+$ states at 19.07 and 19.24 MeV and to the $3^+$ resonance 
at 21.5 MeV.  The reaction rate has been determined by {\it R-}matrix fits to 
the experimental data with uncertainties $\sim 1\%$~\cite{serpicoetal,adahchour}.
More refined theoretical approaches based on modern coupled-channel effective 
field theory lead essentially to the same conclusion~\cite{Lensky}.   

The cross section of $^7{\rm Be}(n,p)^7{\rm Li}$ reaction is extremely large. 
The Maxwellian-averaged cross section at thermal energies 
$\langle \sigma_{\rm np}\rangle_{T}= 3.84 \times 10^4 
\,{\rm barn} $ \cite{Ajzenberg,Koehler} is
the largest thermal cross section known in the light element region.
At the relevant energies for $^7$Be synthesis, 
$E_{\rm Be} \simeq  T_{\rm Be, f} \simeq 50 \,{\rm keV}$, we have 
$\sigma_{\rm np}(E_{\rm Be}) \simeq 9 \, {\rm barn}$; a value quite close to the 
unitarity bound value of 
$\sigma_{\rm max}(E_{\rm Be}) = \pi/(2 \mu E_{\rm Be}) \simeq 15 \,{\rm
barn}$. Therein $\mu$ is the reduced mass of the $^{7}{\rm Be}+n$ system.
The size of this cross section makes it difficult to find comparable
channels for $^7$Be destruction. 

As an alternative process, we consider the reaction 
$^{7}{\rm Be}(n,\alpha)^{4}{\rm He}$ 
for which no experimental data exist in the energy range relevant
for primordial nucleosynthesis. This process normally is included in BBN 
calculations by using the old reaction rate estimate~\cite{Wagoner} to which 
is conservatively assigned a factor 10 uncertainty~\cite{serpicoetal}.  
The process $^{7}{\rm Be}(n,\alpha)^{4}{\rm He}$ is the second most important 
contribution to the $^7{\rm Be}$ rate of destruction, accounting for
$\sim 2.5\%$ of the total.  Nevertheless, due to the large uncertainty
assigned, it provides the dominant contribution to theoretical errors
in the $^{7}{\rm Li}$ abundance evaluations~\cite{serpicoetal}.

To obtain a factor of 2 reduction in the cosmic $^{7}{\rm Li}$ abundance,  
the cross section of the $^{7}{\rm Be}(n,\alpha)^{4}{\rm He}$ reaction need be 
increased $\sim 60$-fold to have
$\sigma_{\rm n\alpha}(E_{\rm Be}) \simeq 1.5  \, 
\sigma_{\rm np}(E_{\rm Be})\sim 15 \,{\rm barn}$; an extremely unlikely 
possibility.  An upper bound on the non-resonant contribution to 
$\sigma_{n\alpha}$ can be obtained by considering the upper limit on the 
Maxwellian-averaged cross section  
$\langle\sigma_{\rm n\alpha}\rangle_T\le 0.1 \; {\rm mbarn}$ 
that was derived~\cite{bassi63} using thermal neutrons. 
Due to parity conservation of strong interactions, the process cannot proceed 
via an $s-$wave collision.  
If we assume a $p-$wave collision, the measured value can be rescaled to 
$T_{\rm Be,f}\sim 50 {\rm keV}$ according to  
$\langle \sigma_{\rm n\alpha} \rangle_T \propto \sqrt T$, obtaining the value
$\langle \sigma_{\rm n\alpha} \rangle_{T_{\rm Be, f}} \le 0.02 \, 
\langle \sigma_{\rm np} \rangle_{T_{\rm Be, f}}$.
That result is much lower than what is required to solve the 
$^{7}{\rm Li}$ problem.  One can question this estimate because it involves
extrapolation over several orders of magnitude. Irrespective of this, 
the process $^{7}{\rm Be}(n,\alpha)^{4}{\rm He}$ will be suppressed at low 
energies with respect to $^{7}{\rm Be}(n,p)^{7}{\rm Li}$ because of 
centrifugal barrier penetration; a quantitative estimate of  which
can be obtained by using the results discussed in the previous section. At low 
energy, 
$\sigma_{\rm n\alpha}/\sigma_{\rm np} \sim T_{1} /T_{0} \sim 2\,\mu\, E \, R^2$
where $T_{0}$ ($T_{1}$) is the transmission coefficient for an $s-$wave ($p-$wave) 
collision in the $^{7}{\rm Be}+n$ entrance channel. 
By considering $E = E_{\rm Be}$, and by taking $R \le 10 \, {\rm fm}$ 
as a conservative upper limit for the entrance channel radius,  we obtain
$\sigma_{\rm n\alpha}(E_{\rm Be}) \le 0.2 \; \sigma_{\rm np}(E_{\rm Be})$;
that is also insufficient to explain the $^7{\rm Li}$ discrepancy.

%
 
Finally, we note that we do not expect a large resonant contribution to the 
$^{7}{\rm Be}(n,\alpha)^{4}{\rm He}$ cross section.
The $^{8}{\rm Be}$ excited states relevant for $^{7}{\rm Be}(n,p)^{7}{\rm Li}$ 
reaction, due to parity conservation, do not decay by $\alpha$-emission.
In summary, in view of experimental and theoretical considerations, 
it appears unlikely that the $^7{\rm Be}+n$ destruction channel is underestimated 
by the large factor required to solve the $^{7}{\rm Li}$ problem.  

\begin{figure}
\par
\begin{center}
\includegraphics[width=7.5cm,angle=0]{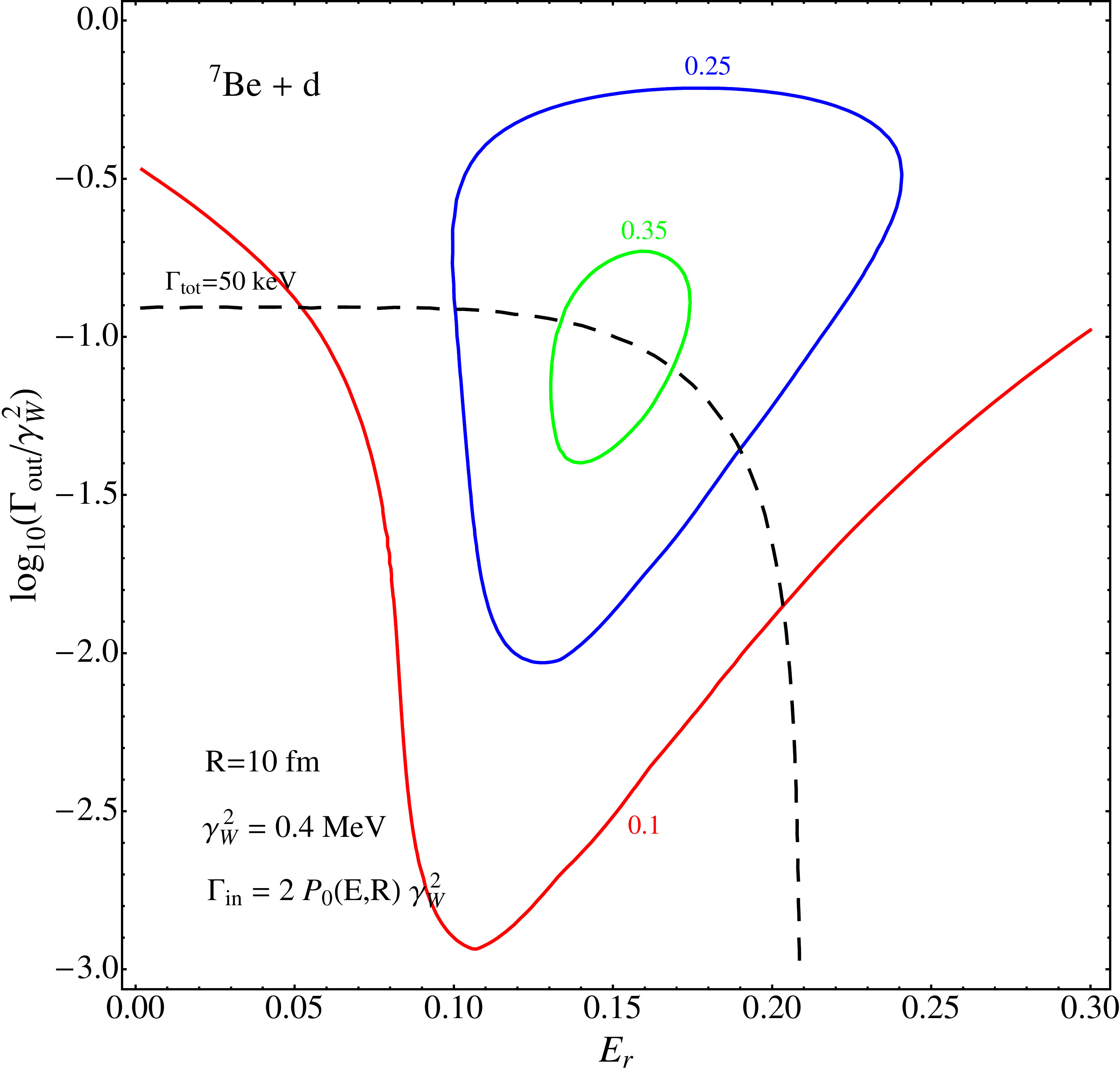}
\includegraphics[width=7.5cm,angle=0]{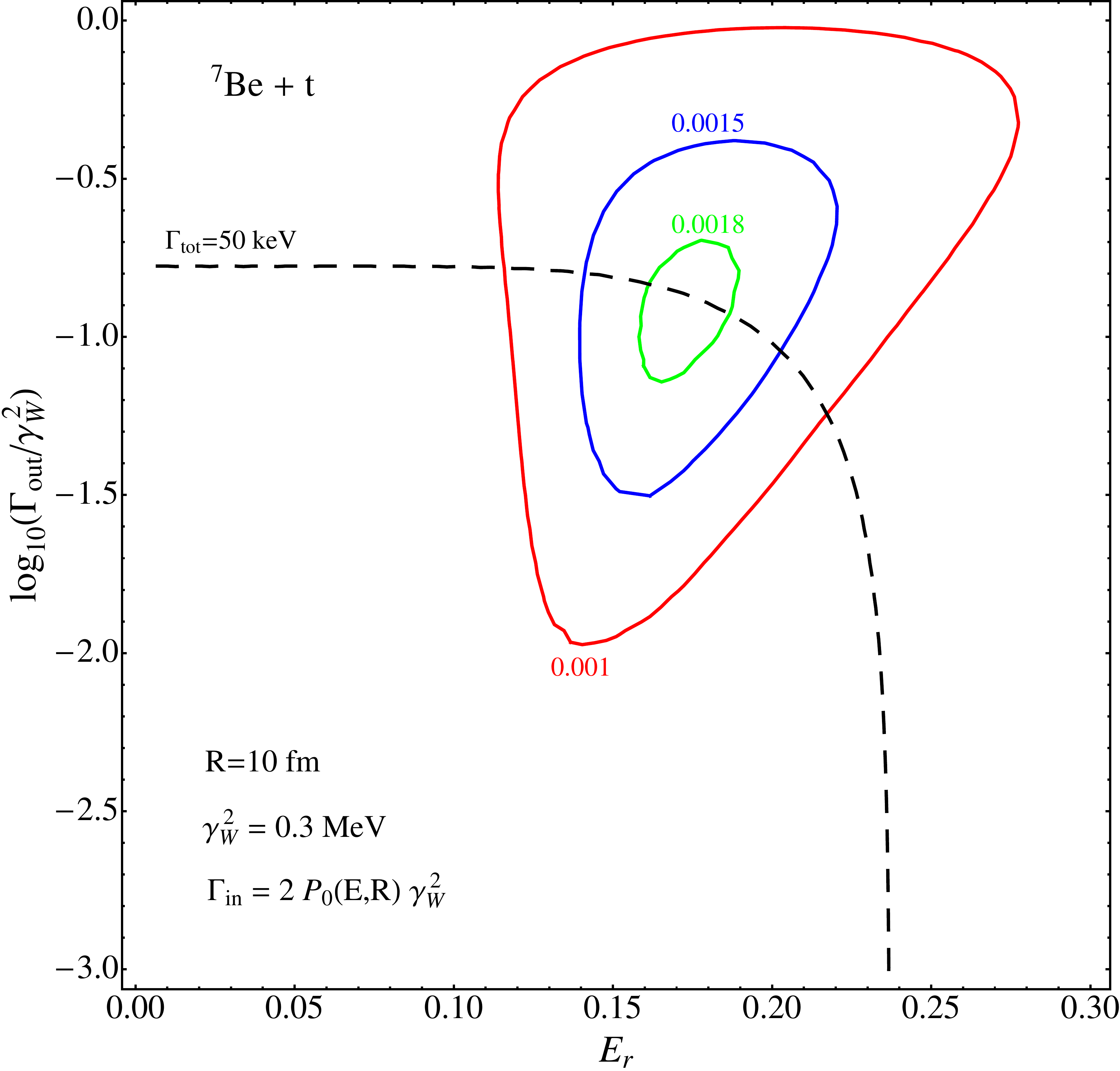}
\includegraphics[width=7.5cm,angle=0]{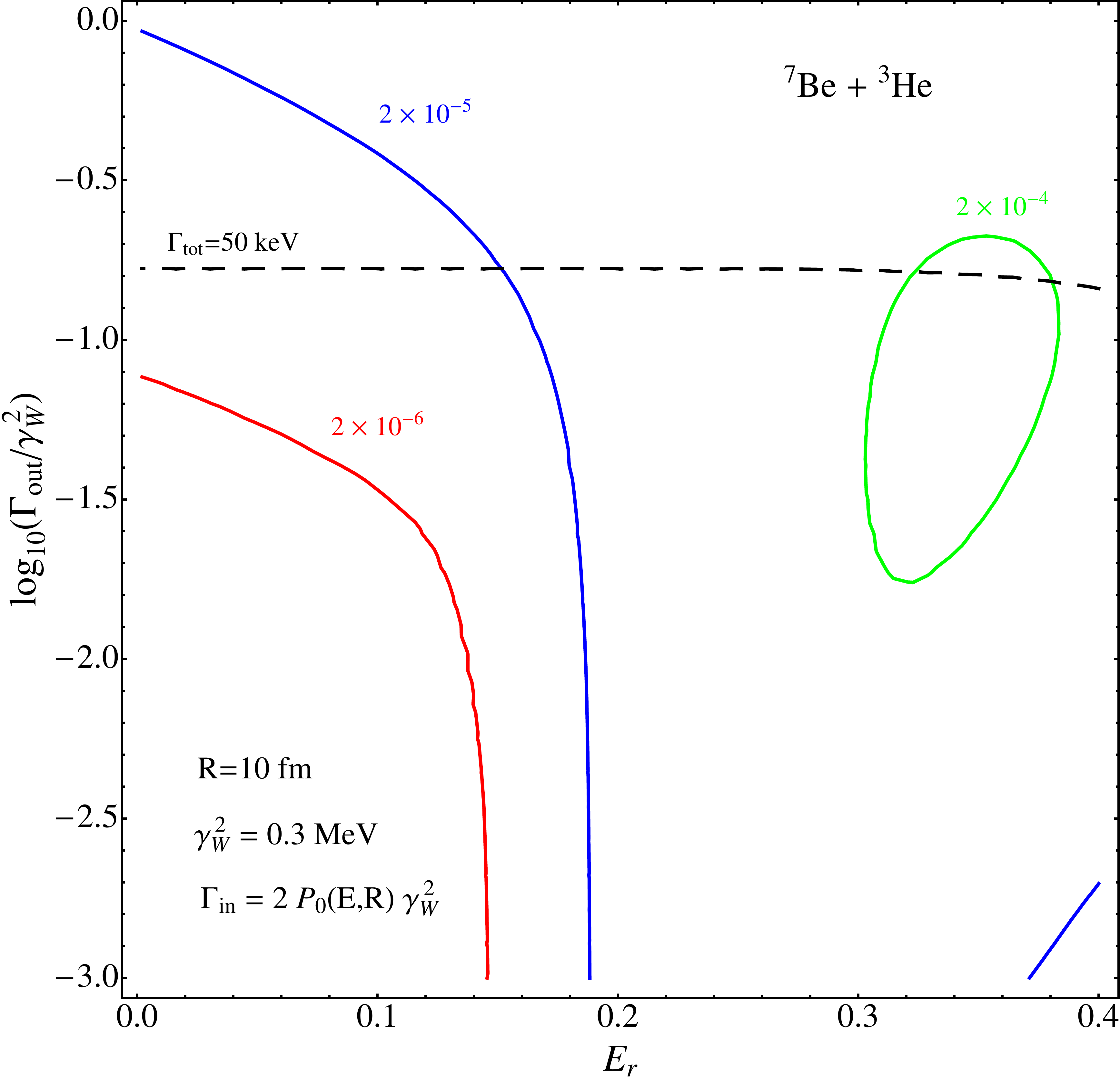}
\includegraphics[width=7.3cm,angle=0]{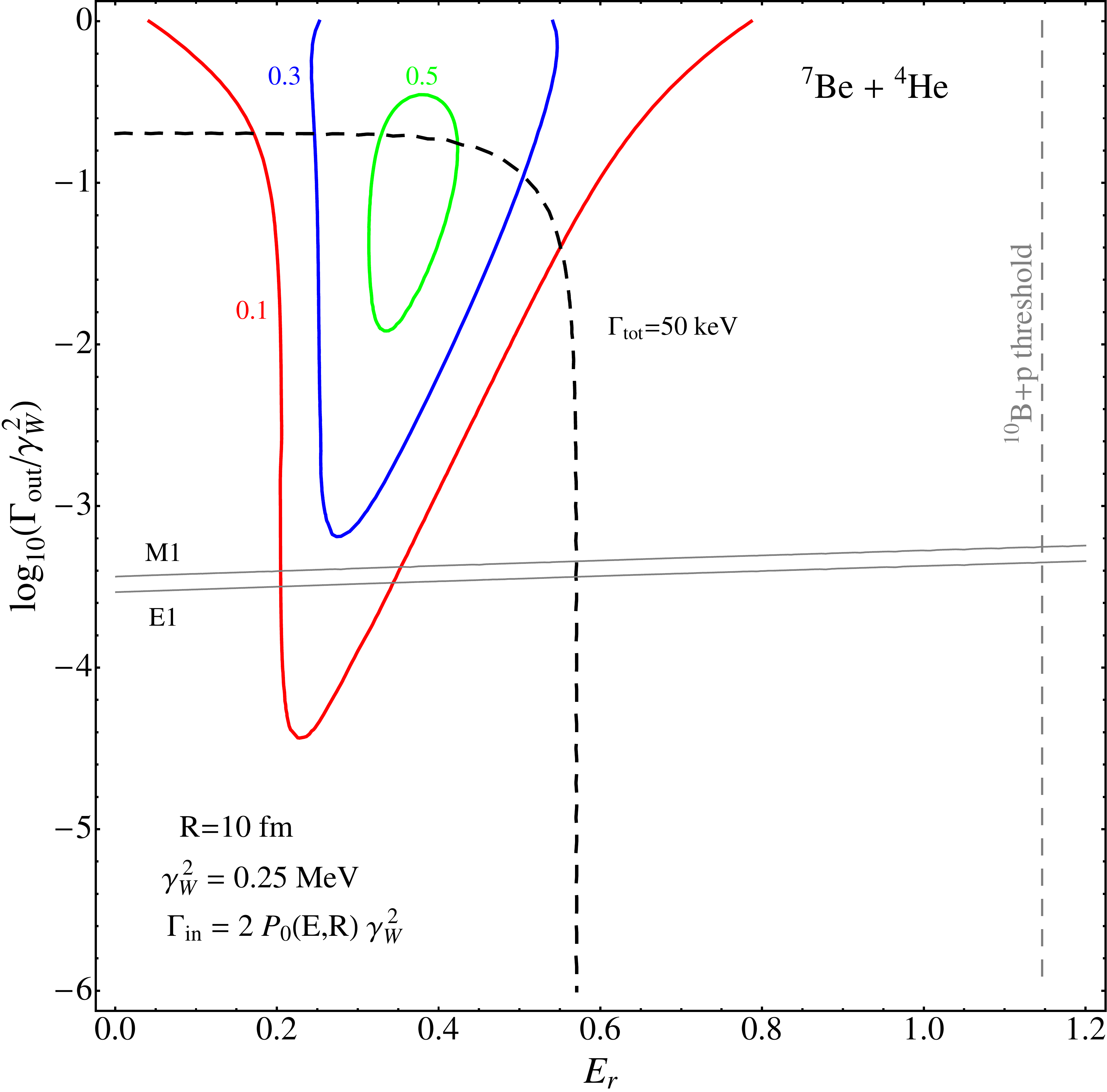}
\end{center}
\par
\vspace{-5mm} \caption{\em {\protect\small   
The coloured 
lines show the fractional reduction of the primordial 
$^7{\rm Li}$ abundances that can be achieved by a resonance in the  
$^{7}{\rm Be}+a$ reaction.  The various panels correspond to 
$a = d,\; t,\; ^3{\rm He}$ and $^4{\rm He}$, respectively, starting from the 
upper-left corner.  The black dashed lines correspond to the condition 
$\Gamma_{\rm tot}(E_r,R) = 50 \,{\rm keV}$, which is the limit for narrow resonance.
The gray solid lines in the lower-right panel correspond to the upper limits for $\Gamma_{\rm out}$
when we assume E1 and M1 electromagnetic transitions to $^{11}{\rm C}$ ground state.}}
\label{Fig3}
\end{figure}

\section{
Other possible \boldmath $^7{\rm Be}$ destruction channels}
\label{Becharged} 

 In standard BBN, the $^7{\rm Be}$ destruction channels involving charged nuclei 
are strongly sub-dominant.  To produce sizeable effects on the $^{7}{\rm Li}$ 
abundance, their efficiency has to be increased by a very large factor. 
This seems possible only if new unknown resonances are found. We discuss this 
possibility by using the Breit-Wigner formalism.

For resonances induced by charged particle reactions at energies below the 
Coulomb barrier, the partial width of the entrance channel $\Gamma_{\rm in}$
varies very rapidly over the resonance region. That is due to the energy dependence of 
$P_{l}(E,R)$, see Eq.(\ref{gammain}). In general, the partial width $\Gamma_{\rm out}$ of 
the exit channel varies more slowly since the energy of the emitted particle is 
increased by an amount equal to the $Q$-value of the reaction. We consider
such a case and so neglect energy dependence of $\Gamma_{\rm out}$ to have 
a cross section form,
\begin{equation}
\sigma_a = \frac{\pi\,\omega\, P_{l}(E,R)}{2 \mu \, E} 
\frac{2\,\xi}{\left[(E-E_{r})/\gamma_{\rm in}^2\right]^2
+\left[2 P_{l}(E,R)+\xi\right]^2/4} ,
\end{equation}
where it has been assumed that 
$\Gamma_{\rm tot} \simeq \Gamma_{\rm in} + \Gamma_{\rm out}$ and 
$\xi = \Gamma_{\rm out} /  \gamma^2_{\rm in}$. Then, for any chosen values 
of $(E_{r},\xi)$ and for any energy $E$, the cross section is an increasing
function of the reduced width $\gamma^{2}_{\rm in}$.  To maximise this cross 
section,
we assume that the reduced width of the entrance channel is equal the Wigner 
limiting width (Eq.~(\ref{gammaW})) that represents
the maximum possible value in the approximation where the interaction potential 
is modelled as a square well of radius $R$. 
%
Moreover, we  assume that the entrance channel is an $s$-wave ($l=0$)
and that the factor $\omega$ has the maximum value allowed by angular momentum 
conservation by  setting
$J_{C^*}=J_{a}+J_{\rm Be}$ in Eq.(\ref{omega}).
Under these assumptions, the cross section of the resonant process is
\begin{equation}
\sigma_a = \frac{\pi\,\omega\, P_{0}(E,R)}{2 \mu \, E} 
\frac{2 \xi}{\left[(E-E_{r})/\gamma^2_{\rm W}(R)\right]^2
+\left[2\,P_{0}(E,R)+\xi\right]^2/4} .
\label{fe}
\end{equation}
It is a function of the two resonance parameters, $(E_{r},\xi)$, and of the 
entrance channel radius $R$. 

We have applied Eq.~(\ref{fe}) to a generic $^7$Be destruction channel involving 
charged nuclei and then determined the  effect on the $^{7}{\rm Li}$ abundance by 
using Eq.~(\ref{inverse2}).  The thermally averaged cross section 
$\langle \sigma_{a} v \rangle_{T}$ has been evaluated numerically
without using the narrow-resonance approximation.
%
Our results are shown in Fig.~\ref{Fig3} as a function of the resonance 
parameters $(E_{r},\xi)$.  The 'coloured'
lines represent the iso-contours for the $^7$Li abundance,
\begin{equation}
\delta Y_{\rm Li} = 1- \frac{Y_{\rm Li}}{\overline{Y}_{\rm Li}} .
\end{equation}
The various panels correspond to the processes $^{7}{\rm Be}+a$ with 
$a = d,\; t,\; ^3{\rm He}$ and $^4{\rm He}$ respectively, starting from the 
upper left corner.\footnote{
We do not consider the $^{7}{\rm Be}+p$ entrance channel since this is known to be subdominant, see e.g. \cite{Chakraborty:2010zj}, and 
it is well studied at low energies due to its importance for solar neutrino production.}
 In our calculations, we assumed the entrance channel 
radius to be $R=10\,{\rm fm}$. That
is quite a large value considering the radii of the involved nuclei
but it has been chosen
to provide a conservative upper estimate of the resonance effects.

It is important to note that there is a maximum achievable reduction 
$(\delta Y_{\rm Li})_{\rm max}$ of the $^{7}{\rm Li}$ 
abundance for each reaction channel considered; a point not discussed 
in previous analyses~\cite{Chakraborty:2010zj, Cyburt:2009cf}. Those
analyses included the effects of Coulomb barrier penetration {\em a posteriori}
and used the narrow-resonance approximation to calculate 
$\langle \sigma_{a} v \rangle_{T}$.
The use of the narrow-resonance approximation outside its regime of validity
can lead to severe overestimation of the resonance effects. 
Indeed,  using that assumption one has 
$\langle \sigma_a v \rangle_{T} \sim \Gamma_{\rm eff} \; (\mu T)^{-3/2} 
\exp(- E_{r}/T)$ where the effective resonance width is defined as 
$\Gamma_{\rm eff} \equiv (\Gamma_{\rm in} \Gamma_{\rm out}) / \Gamma_{\rm tot}$.  
This expression does not predict any upper limit for 
$\langle \sigma_a v \rangle_T$ as a function of $\Gamma_{\rm eff}$.
But a limit should exist. That can be understood by considering 
$\sigma_a \le (\pi \omega) / (2\mu E)$ 
for any possible choice of the resonance parameters as per Eq.(\ref{BWA}). 
In fact, the correct scaling for broad resonances is given by 
$\langle \sigma_a v \rangle_{T} \sim (\Gamma_{\rm eff} / 
\Gamma_{\rm tot}) \; \mu^{-1/2} \;T^{-3/2}$
where the factor $\Gamma_{\rm eff}/\Gamma_{\rm tot}  = 
(\Gamma_{\rm in} \Gamma_{\rm out}) / \Gamma_{\rm tot}^2 $,
cannot be larger than one.
In Fig.~\ref{Fig3}, we display with the `black'
 dashed line the result found using the condition 
$\Gamma_{\rm tot}(E_r,R) =\Gamma_{\rm in}(E_r,R) 
+ \Gamma_{\rm out} = 50 \,{\rm keV}$. 
The turning point of the line corresponds to the situation 
$\Gamma_{\rm in}(E_{\rm r},R) \sim \Gamma_{\rm out}$
and it is localised in the region where $\xi \sim 2\,P_{0}(E_{\rm r},R)$. 
Results with a narrow-resonance approximation can only be compared with
those shown  in the lower left corner of the plots 
where $\Gamma_{\rm tot}(E_{r},R) \ll E_{r}$, and the $^7$Li reduction
typically is negligible. 

\begin{figure}[t]
\par
\begin{center}
\includegraphics[width=7.5cm,angle=0]{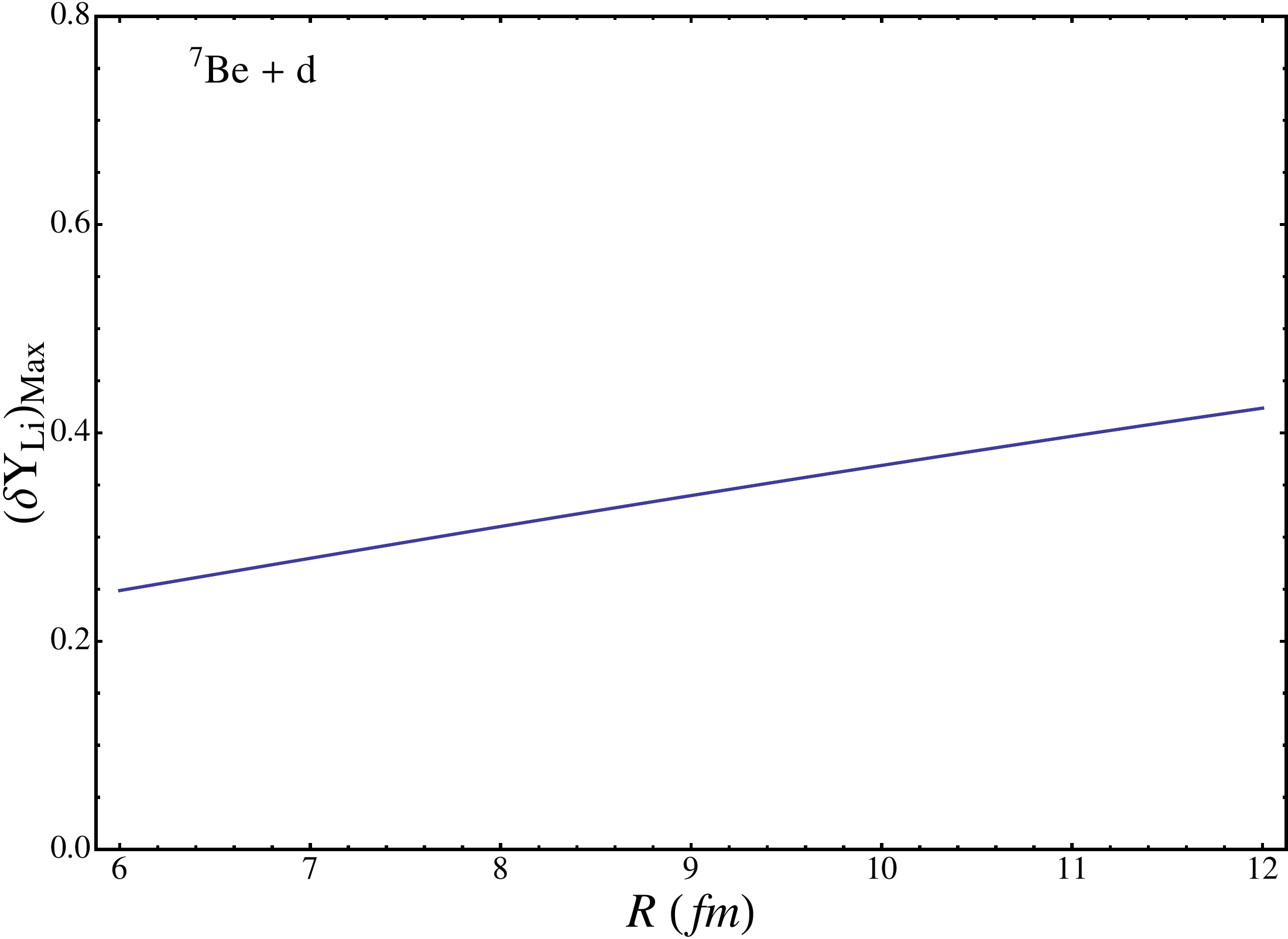}
\includegraphics[width=7.5cm,angle=0]{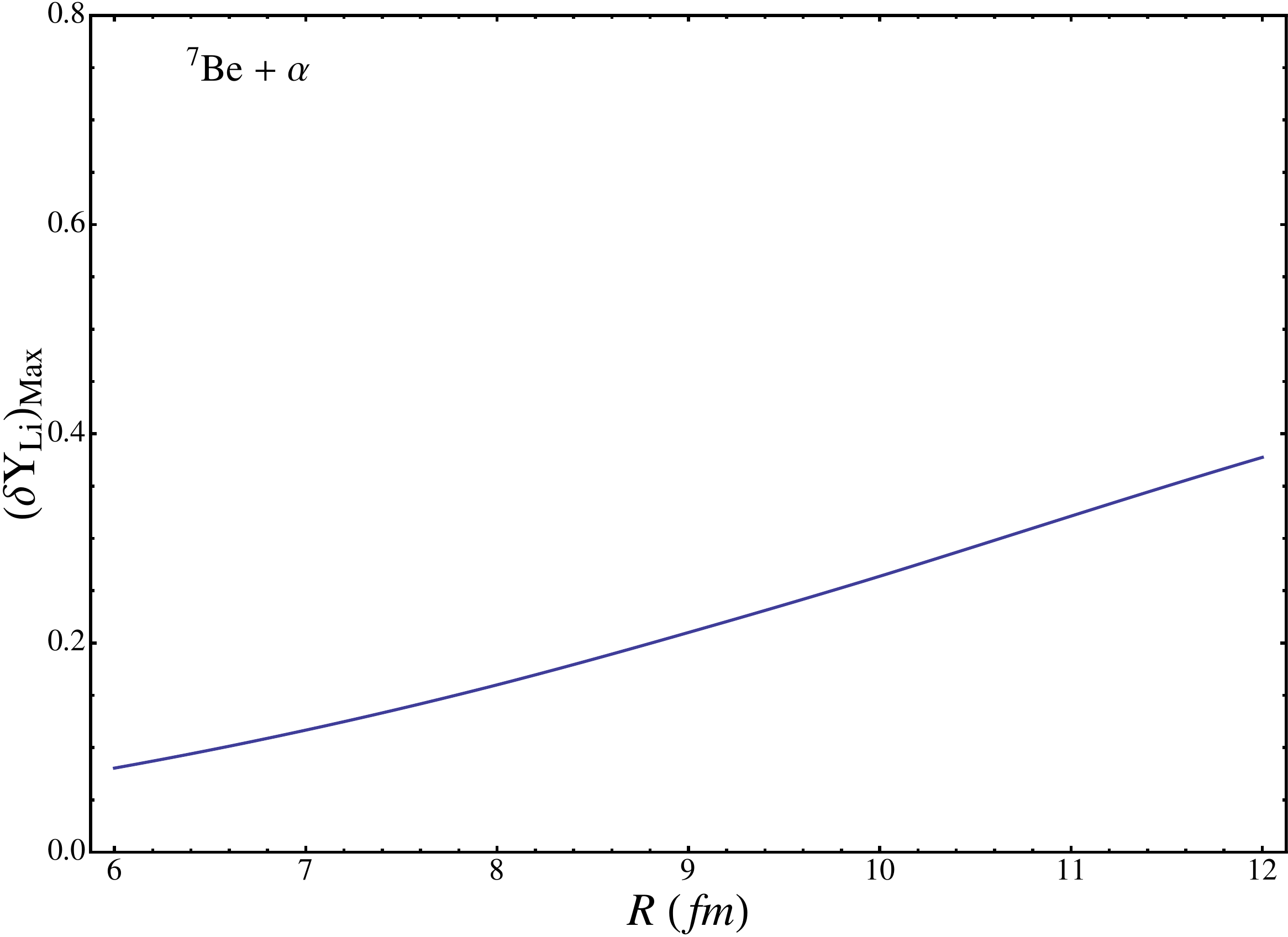}
\end{center}
\par
\vspace{-5mm} \caption{\em {\protect\small  
The maximal fractional reduction $(\delta Y_{\rm Li})_{\rm max}$ of the 
$^7{\rm Li}$ abundance that can be achieved by resonances in $^7{\rm Be}+d$ and $^7{\rm Be}+\alpha$ channels as a function 
of the assumed entrance channel radius $R$. In the case of the $^7{\rm Be}+\alpha$ reaction, we considered
the upper limit $\Gamma_{\rm out} \le 10 \; {\rm W.u.}$ for a magnetic (M1) dipole transitions 
to the $^{11}{\rm C}$ ground state.}}
\label{Fig4}
\end{figure}

The results that we have obtained for each specific reaction channel are 
now outlined sequentially:

\paragraph{$^7$Be + d:} 
With this initial channel,
the maximum achievable effect is a $\sim 40\%$ reduction of primordial 
$^{7}{\rm Li}$ abundance.  This reduction could substantially alleviate the 
discrepancy between theoretical predictions and observational data
which differ by a factor $\sim 3$ in the standard scenario. The maximal effect
is obtained for a resonance energy $E_{r}\sim 150 \; {\rm keV}$  with a 
total width $\Gamma_{\rm tot}(E_{r},R) \sim 45 \, {\rm keV}$ and partial 
widths approximately equal to $\Gamma_{\rm out} \sim 35 \,{\rm keV}$ 
and $\Gamma_{\rm in}(E_{r},R) \sim 10 \,{\rm keV}$. 
The dependence of the maximum reduction $(\delta Y_{\rm Li})_{\rm max}$
on the assumed entrance channel radius $R$ is shown\footnote{The resonance parameters that maximise
the $^7$Li suppression are slightly dependent on the assumed radius.} in 
Fig.~\ref{Fig4} in which $(\delta Y_{\rm Li})_{\rm max}$ increases with $R$.
That is expected given that the partial width of the entrance channel is 
determined primarily by Coulomb barrier penetration as in Eq.~(\ref{gammain}). 
So quite large values for $R$ are needed to solve the cosmic $^7$Li problem. 
Even if these are much larger than the sum of the radii of the involved nuclei,
2.65 fm and 2.14 fm for $^7$Be and the deuteron respectively, 
they cannot be excluded in view of the large uncertainties in the approach.
Our results basically coincides with those found~\cite{Cyburt:2009cf} 
using a different approach.
Note that there is an excited state in $^{9}{\rm B}$ at 16.71 MeV. It lies just 220 keV above the  $^7$Be + d  threshold
and it decays by gamma and particle (proton or $^3$He) emission.
However, this state has been  very recently ruled out as a solution of the cosmic $^{7}{\rm Li}$  problem \cite{Kirsebom}. 
A non negligible suppression would, therefore, require the existence of a new (not yet 
discovered)  excited state of $^{9}{\rm B}$ around $E_r\sim 150\,{\rm  keV}$. 
In \cite{O'Malley}, this possibility was studied by using the $^2{\rm H}(^{7}{\rm Be}, d) ^{7}{\rm Be}$
reaction. The data show no evidence for new resonances and allow to set an upper limit 
on the total resonance width at the level of $\sim 1\, {\rm keV}$.
Finally, we observe that the process 
$^7{\rm Be}+d \to{} ^6{\rm Li} +{} ^{3}{\rm He}$ could enhance $^{6}{\rm Li}$ production in the early universe. 
However, the effect on the {\it final} $^{6}{\rm Li}$ abundance appears steadly too small to explain 
the observations of Ref.\cite{Asplund2006}.

\paragraph{$^7$Be + t:} 
With this initial channel,
the maximum achievable effect is a $\sim 0.2\%$ reduction of primordial 
$^{7}{\rm Li}$ abundance. The existence of a resonance in this channel 
cannot solve the cosmic $^{7}{\rm Li}$ problem, since to produce a significant 
$^7$Li reduction, the $^7{\rm Be}$ destruction rate due to the $^{7}{\rm Be}+t$ 
reaction should be comparable to that form  $^{7}{\rm Be}+n$ processes.
Clearly that is impossible because:
\begin{enumerate}
\item
neutrons are more abundant than tritons at the relevant temperature 
$T_{\rm Be}\sim 50\, {\rm keV}$ as is seen in Fig.~\ref{Fig1}
\item
the cross section of $^7{\rm Be}(n,p)^7{\rm Li}$ is close to the unitarity 
bound while the $^7{\rm Be}$+$t$ collisions are suppressed by Coulomb 
repulsion. 
\end{enumerate}
Conclusions reached by others~\cite{Chakraborty:2010zj} differ from
ours. Theirs are artifacts from using of the narrow-resonance approximation 
outside of its regime of application.

\paragraph{$^7$Be + $^3$He:} 
With this channel, the maximum achievable effect is a $\sim 10^{-4}$ reduction 
in the abundance of primordial $^{7}{\rm Li}$. Again then, a resonance in 
this channel cannot solve the cosmic $^{7}{\rm Li}$ problem. The small
effect is due to strong Coulomb repulsion suppressing the partial width of the 
entrance channel ($\Gamma_{\rm in}(E_r, R)$). From our calculations, 
we obtain $\Gamma_{\rm in}(E_r, R) \sim 0.2 \, {\rm meV}$  and 
$\Gamma_{\rm in}(E_r, R) \sim 7 \, {\rm eV}$ for $E_r \simeq 100\, {\rm keV}$ and $E_r \simeq 200\, {\rm keV}$ respectively; 
values that are much smaller than required to obtain non negligible effects.

\paragraph{$^7$Be + $\alpha$:} 
With this channel, the maximum achievable effect is, in principle, a $\sim 55\%$ reduction 
of primordial $^{7}{\rm Li}$ abundance.
This is obtained for a resonance with a relatively large centroid energy 
$E_{r}\sim 360 \; {\rm keV}$, with a total width 
$\Gamma_{\rm tot}(E_{r},R) \sim 21 \, {\rm keV}$ and with partial widths 
$\Gamma_{\rm out} \sim 19 \,{\rm keV}$ and 
$\Gamma_{\rm in}(E_{r},R) \sim 1.5 \,{\rm keV}$. 
The strong suppression of the cross section due to Coulomb repulsion in
this case is compensated by the fact that the $\alpha$ nuclei are $\sim 10^6$ 
times more abundant than neutrons when the temperature of the universe
falls below $\sim 70\,{\rm keV}$. However, one should note that
for $E_r \le 1.15~{\rm MeV}$ there are no particle exit channels for the coumpound
$^{11}{\rm C}$ nucleus.  As a consequence, the only possible transition is the electromagnetic one 
whose width is expected to be smaller than $\sim 100\,{\rm eV}$. In Fig.~\ref{Fig3}, 
we show with the gray solid lines the recommended upper limits for 
the width of electric (E1) and magnetic (M1) dipole transitions given by \cite{endt93}.
These corresponds to 0.5 Weisskopf units (W.u.) and 10 W.u. respectively and have been  
calculated by assuming a $\gamma-$transition to the $^{11}{\rm C}$ ground state.\footnote{The $\gamma-$ray transitions
corresponding to higher multipolarities are suppressed with respect to M1 and E1 by a factor 30 or more.}
If we consider an $s-$wave collision, 
the quantum numbers of the compound $^{11}{\rm C}$ nucleus only allow M1 radiation to be emitted. 
Taking the corresponding limit into account, one obtains at most a $\sim 25\%$ reduction 
of the $^7{\rm Li}$ abundance for $E_{r}\sim 270 \; {\rm keV}$, with a total width 
$\Gamma_{\rm tot}(E_{r},R) \sim 160 \, {\rm eV}$ and partial widths 
$\Gamma_{\rm out} \sim 100 \,{\rm eV}$ and 
$\Gamma_{\rm in}(E_{r},R) \sim 60 \,{\rm eV}$.\footnote{In Fig.~9 of 
Ref.~\cite{Chakraborty:2010zj}, it was shown that a narrow resonance at
an energy  $E\le 100 \, {\rm keV}$ and effective width 
$\Gamma_{\rm eff}\sim 10\,{\rm meV}$ can substantially reduce the $^7$Li 
abundance.  According to our analysis, however, this region of 
parameter space is unphysical since, if we take into account correctly the effects
of Coulomb repulsion, we obtain $\Gamma_{\rm in}(E_r,R)\le 1\, {\rm meV} $ for 
$E_r \le 100\, {\rm keV}$.}
The dependence of the maximal achievable reduction $(\delta Y_{\rm Li})_{\rm max}$ from the entrance channel radius $R$
is explored in the right panel of  Fig. \ref{Fig4}. 
We note that the existence of a resonance with these parameters would imply a non negligible counting rate
in $^{7}{\rm Be}+ \alpha$ experiments. By following \cite{Iliadis}, we estimate that a $^{7}{\rm Be}$ beam with an intensity 
$\sim 5 \times 10^4 \, ^7{\rm Be/s}$ (i.e. comparable to that used in \cite{O'Malley}) would produce $\sim 50$ events/day 
with the emission of a $\sim 7.8~{\rm MeV}$  gamma ray (or a gamma ray cascade) in a thick $^{4}{\rm He}$ target. 
Such a rate would be measurable in an underground laboratory~\cite{carlo}.

The possibility of a missing resonance in the $^{7}{\rm Be}+\alpha$ channel is particularly 
intriguing theoretically. We have calculated the spectrum of $^{11}$C 
using a coupled-channel model for the $n$-$^{10}$C system, with coupling 
involving the excited quadrupole state of $^{10}$C. A multi-channel
algebraic scattering (MCAS) was used with which account is made of
constraints imposed by the Pauli principle on single-particle dynamics
besides coupling interactions to the collective excitations of the  
$^{10}$C states~\cite{Amos03}. Bound and resonant low-energy spectra
of light nuclei have been analysed systematically with this 
approach~\cite{Can05,Can06} and
in particular for carbon isotopes~\cite{Sve05,Kar08,Amos12}. 
In Table~\ref{C11-spect} a comparison is given between the calculated
spectrum with the observed levels of $^{11}{\rm C}$~\cite{TUNL-C11}. 
Clearly there is a one to one correspondence except for a $\frac{1}{2}^-$ 
state predicted at 6.885 MeV. 
That excitation energy  lies relatively close to entrance of $^{7}{\rm Be}+{}^{4}{\rm He}$ channel
which is 7.543 MeV above the $^{11}$C ground state and would require a $d-$wave collision (or a coupled-channel transition to the 
$^{7}{\rm Be}$ first excited state) to ensure angular momentum and parity conservation.
We remark that the existence of a new state for $^{11}{\rm C}$  
would imply the existence of  a corresponding state for the mirror $^{11}{\rm B}$ nucleus
with comparable energy and width. $^{11}{\rm B}$ is stable and well studied experimentally
by using photon \cite{1980Moreh} and electron scattering \cite{1975Kan} reactions.
At present, there is no evidence for such a state as can be seen by comparing the
observed levels of $^{11}{\rm C}$ and those of $^{11}{\rm B}$~\cite{TUNL-C11} .
We note, moreover, that the $^{11}{\rm B}$ levels have, at most, a few eV energy widths
which are considerably lower that what required to suppress the $^{7}{\rm Li}$ abundance.

\begin{table}[t]
\begin{center}
\begin{tabular}{ccc}
$J^P$ & Nuclear data & MCAS levels\\
\hline
 3/2$^-$  & 0.00 & 0.00 \\
 1/2$^-$  & 2.000 & 2.915 \\
 5/2$^-$  & 4.3188 & 3.225 \\
 3/2$^-$  & 4.8042 & 3.303 \\
 1/2$^+$  & 6.3392 & 8.373 \\
 7/2$^-$  & 6.4782 & 5.768 \\
 5/2$^+$  & 6.9048 & 7.781\\
 1/2$^-$  &   ?    & 6.885 \\
 3/2$^+$  & 7.4997 & 11.059\\
 3/2$^-$  & 8.1045 & 7.332\\
 5/2$^-$  & 8.420  & 9.689\\
 7/2$^+$  & 8.655  & 10.343\\
 5/2$^+$  & 8.699 &  10.698\\
 5/2$^+$  & 9.20 & 11.868 \\
 3/2$^-$  & 9.65 & 11.253 \\
 5/2$^-$  & 9.78 &  12.802\\
 7/2$^-$  & 9.97 & 9.022\\
\end{tabular}
\end{center}
\caption{\label{C11-spect} \em {\protect\small  
Spectra of $^{11}C$. The data are taken from Ref.~\cite{TUNL-C11} while
calculated values have been obtained with the MCAS formalism~\cite{Amos03}. 
Potential parameters defining the MCAS coupled-channel interactions
have not been sought for the optimal reproduction of levels, 
but only to check for possible missing resonances.}}
\end{table}

\section{Conclusions}
\label{Conclusions}

We have investigated the possibility that the  cosmic $^7{\rm Li}$ problem 
originates from incorrect assumptions about the nuclear reaction cross 
sections relevant for BBN.  To do so, we introduced an efficient method to 
calculate the changes in the $^{7}{\rm Li}$ abundance produced by an 
arbitrary (temperature dependent) modification of the nuclear reaction rates. 
Then, taking into account that $^7{\rm Li}$ is mainly produced through 
$^{7}{\rm Be}$, we used this method to assess whether it is possible to 
increase the total $^{7}{\rm Be}$ destruction rate to the level required to 
solve (or alleviate) the cosmic $^{7}{\rm Li}$ puzzle. 
Given present experimental and theoretical constraints, it is
unlikely that the $^7{\rm Be}+n$ destruction rate is underestimated by the 
$\sim 2.5$ factor required to solve the cosmic $^{7}{\rm Li}$ problem.

On the basis of very general nuclear physics considerations, we have shown 
that the only destruction channels that could have a non negligible
impact on the $^7{\rm Li}$ abundance are $^{7}{\rm Be}+d$ and 
$^{7}{\rm Be} +{} \alpha$.  Our results suggest that it is unrealistic 
to consider new resonances in $^{7}{\rm Be}+t$ and 
$^{7}{\rm Be} +{} ^3{\rm He}$ channels to solve the $^7{\rm Li}$ 
problem.  With the other two channels, new resonances must exist at 
specific energies and with suitable resonance widths. 
Postulating a resonance in the $^{7}{\rm Be}+d$ reaction at an
energy $E_{r}\sim 150 \; {\rm keV}$, with a total width  
$\Gamma_{\rm tot}(E_{r},R) \sim 45 \, {\rm keV}$, and partial widths 
$\Gamma_{\rm out} \sim 35 \,{\rm keV}$ and 
$\Gamma_{\rm in}(E_{r},R) \sim 10 \,{\rm keV}$ 
gave a $\sim 40\%$  reduction in the $^{7}{\rm Li}$ abundance.
However, recent experimental results have excluded unknown resonances
with these properties.
A smaller suppression of $\sim 25\%$ was obtained by assuming a resonance
in the $^{7}{\rm Be}+{} \alpha$ channel with energy 
$E_{r}\sim 270 \; {\rm keV}$, with a total width 
$\Gamma_{\rm tot}(E_{r},R) \sim 160 \, {\rm eV}$ 
and partial widths $\Gamma_{\rm out} \sim 100 \,{\rm eV}$ and 
$\Gamma_{\rm in}(E_{r},R) \sim 60 \,{\rm eV}$.
These  results are the maximal achievable reductions of the 
$^{7}{\rm Li}$ abundance, since they were obtained assuming that the 
resonance width of the entrance channel has the largest value allowed in the 
presence of Coulomb repulsion, and by scanning the (allowed) space of the other resonance 
parameters.  Also, we considered a relatively large value for the 
entrance channel radii, $R=10\,{\rm fm}$.

In summary, 
the present study reduces significantly the space for a nuclear physics solution of the cosmic $^{7}{\rm Li}$ problem.
Even a partial solution would require an extremely favorable combination 
in the character of still undetected resonances in $^{7}{\rm Be}+\alpha$ channel and such  a possibility
could be excluded by  new experimental efforts.

\acknowledgments{We are grateful to K.~Amos for critical reading of the manuscript. We thank O.S. Kirsebom and the participants of 
``Lithium in the cosmos'' for useful discussions. We thank M.~Pospelov for pointing out the importance of 
missing low-energy particle exit channels for the $^7{\rm Be}+\alpha$ reaction. 

\newpage

\section*{\sf  References}
\def\refname{

\end{document}